\documentclass[12pt]{article}
\pdfoutput=1
\usepackage{euscript,amsmath,amsthm,amsfonts,amssymb}
\usepackage{mathtools}
\usepackage{subfig}
\usepackage{float}
\usepackage[margin=1in]{geometry}
\usepackage{graphicx}
\usepackage{outlines}
\usepackage[dvipsnames]{xcolor}
\usepackage{jheppub}
\newcommand{\ket}[1]{|{#1}\rangle}

\newcommand{\bra}[1]{\langle{#1}|}

\newcommand{\be}{\begin{equation}}
\newcommand{\ee}{\end{equation}}
\newcommand{\bea}{\begin{eqnarray}}
\newcommand{\eea}{\end{eqnarray}}

\title{\boldmath Level-rank duality of $SU(2)_k$ Chern-Simons theory, and of hypergraph and magic states}
\author{Howard J. Schnitzer}
\affiliation{Martin Fisher School of Physics, Brandeis University, Waltham, Massachusetts 02453, USA}
\preprint{BRX-TH-6677}
\emailAdd{schnitzr@brandeis.edu}
\abstract{The level-rank duality of $SU(2)_{k}$ Chern-Simons theory is discussed, and applied to graph, hypergraph, and magic states.

}
\begin{document}
\maketitle
\flushbottom

\section{Introduction}
Level-rank duality of Chern-Simons theory and WZW theories has a long history, \cite{1990PhLB..244..235N,NACULICH1990687,proc,NACULICH1990417,MLAWER1991863,Naculich:1992uf} with many applications. It generally involves transposition of the Young tableaux of the representations for a pair of theories, linked by the duality. However, in general level-rank duality is not a one-to-one map between the tableaux of the two theories. For example, $SU(2)_3$ has four states, while $SU(3)_2$ has six. However, using cominimal equivalent or simple current maps relating states, \cite{MLAWER1991863} a one-to-one map of the representations is possible. Note that $SU(2)_3/ \mathbb{Z}_3$ and $SU(3)_2/ \mathbb{Z}_2$ have just two states related one-to-one by transposition of the respective Young tableaux. The generalization of this theme is central to this paper.

It is the purpose of this paper to discuss the level-rank duality of $SU(2)_k$ and its relationship to $SU(k)_2$. The generalized Pauli group can be defined for each side of the duality, which can then be used to construct graph and hypergraph states for the level-rank pairs. Since a sub-set of hypergraph states describe magic states, level-rank duals of magic states also results.

\section{Level-rank dualities of $SU(2)_k$}
The integral representations of $SU(2)_k$ Chern-Simons theory can be described by a single-row Young tableau with $0\leq k$ boxes, while that of $SU(k)_2$ is described by a Young tableau with two columns, $l_1$ and $l_2$, where $l_1\geq l_2$, and $l_1$ ranges from 0 to $k-1$. Thus, $SU(2)_k$ has $k+1$ states, while $SU(k)_2$ has $\frac{1}{2}k(k+1)$ states. Therefore a transpose map between the representations of $SU(2)_k$ and $\widetilde{SU}(k)_2$ is not one-to-one, where $\widetilde{SU}$ indicates the transpose of the Young tableau. However, making use of the orbits of cominimal maps or equivalently simple current maps, \cite{MLAWER1991863} one can obtain level-rank dual maps which are one-to-one for the Young tableaux. The two cases that we focus on are
\be\label{eq:21}
SU(2)_{2k}=\tilde{SU}(2k)_2/\mathbb{Z}_k
\ee
which has $2k+1$ states, and
\be\label{eq:22}
SU(2)_{2k+1}/\mathbb{Z}_2=\tilde{SU}(2k+1)_2/\mathbb{Z}_{2k+1}
\ee
which has $k+1$ states. In the next subsection we consider the generalized Pauli group for these two cases.

\subsection{Generalized Pauli group}\label{sec:2b}

The qudit Pauli group for a $d$-dimensional system, for both $d$ odd and even, is described by Farinholt \cite{Farinholt_2014}. One defines the operators
\be
x=\sum_{x \in \mathbb{Z}_\alpha}\ket{x+1}\bra{x}
\ee
and
\be
Z=\sum_{x\in \mathbb{Z}_\alpha} \omega^x\ket{x}\bra{x}
\ee
where $\omega=\exp\left(\frac{2\pi i}{d}\right)$ is a primitive root of unity. We focus on $d$-odd for simplicity for the dual pairs \eqref{eq:21} and \eqref{eq:22}. For the dual pairs of \eqref{eq:21}, $d$ is odd for any integer $k$. However, for \eqref{eq:22}, $d$ odd requires $k$ even.

With these restrictions, the operators $X$ and $Z$ are defined for the level-rank dual pairs \eqref{eq:21} and \eqref{eq:22}. Given the level-rank dual pairs \eqref{eq:21} and \eqref{eq:22}, and the restriction to $d$ odd, identify $X=\tilde{X}$ and $Z=\tilde{Z}$ for the Pauli operators of both sides of \eqref{eq:21} and \eqref{eq:22}. Thus the dual pairs of $X=\tilde{X}$ and $Z=\tilde{Z}$ enables one to define level-rank dual pairs of graphs and hypergraph states, which we consider in the next sub-section.

\subsection{Graph and hypergraph states}
\paragraph{Graph states}

There are many equivalent constructions of graph states \cite{Steinhoff_2017,Englbrecht_2020,liu2020manybody,Rossi_2013,2006quant.ph..2096H}. We follow arxiv:1612.06418 for a definition of qudit graph states. The multigraph is $G = (V,E)$, with vertices $V$ and edges $E$, where an edge has multiplicity $m_e \in \mathbb{Z}_d$. To $G$ associate a state $\ket{G}$ such that to each vertex $i \in V$, there is a local state
\begin{equation}
    \ket{+} = \ket{p_0} = \frac{1}{\sqrt{d}} \sum_{q = 0}^{d-1} \ket{q}
\end{equation}
Define
\begin{align}
    S^* \ket{0} &= \frac{1}{\sqrt{d}} \sum_{q = 0}^{d-1} \ket{q} \nonumber \\
    &= \ket{+} = \ket{p_0}.
\end{align}

To each edge $e = \{i,j\}$ apply the unitary
\begin{equation}\label{7}
    Z_e^{m_e} = \sum_{q_i = 0}^{d-1} \ket{q_i} \bra{q_i} \otimes \left(Z_j^{m_e}\right)^{q_i}
\end{equation}
to the state
\begin{equation}\label{8}
    \ket{+}^V = \bigotimes_{i \in V} \ket{+}_i
\end{equation}

The graph state is
\begin{align}\label{9}
    \ket{G} &= \prod_{e \in E} Z_e^{m_e} \ket{+}^V \\
        &= \prod_{e \in E} Z_e^{m_e} \bigotimes_{i \in V} \ket{+}_i\label{10}
\end{align}
Every stabilizer state is LC equivalent to a graph state, while the Clifford group enables conversion between different multigraphs \cite{Steinhoff_2017,Englbrecht_2020,liu2020manybody,Rossi_2013,2006quant.ph..2096H}.

\paragraph{Hypergraph states}

We again follow arxiv:1612.06418 for the construction of qudit multi-hypergraph states. Given a multi-hypergraph $H = (V,E)$, associate a quantum state $\ket{H}$, with $m_e \in \mathbb{Z}_d$ the multiplicity of the hyperedge $e$. To each vertex $i \in V$, associate a local state
\begin{align}
    \ket{+} &= \frac{1}{\sqrt{d}} \sum_{q=0}^{d-1} \ket{q}
\end{align}
To each hyperedge $e \in E$, with multiplicity $m_e$, apply the controlled unitary $Z_e^{m_e}$ to the state
\begin{equation}
    \ket{+}^V = \bigotimes_{i \in V} \ket{+}_i
\end{equation}
The hypergraph state is
\begin{align}\label{13}
    \ket{H} = \prod_{e \in E} Z_e^{m_e} \ket{+}^V
\end{align}
and the elementary hypergraph state is
\begin{align}\label{14}
    \ket{H} = \sum_{q=0}^{d-1} \ket{q} \bra{q} \otimes \left( Z_{e \backslash \{1\}}^{m_e} \right)^q \ket{+}^V
\end{align}
For $d$ prime, all $n$-elementary hypergraph states are equivalent under SLOCC.

Hypergraph and graph states admit a representation in terms of Boolean functions,
\begin{equation}
    \ket{H} = \sum_{q = 0}^{d-1} \omega^{f(q)} \ket{q}
\end{equation}
with $f: \mathbb{Z}_d^n \rightarrow \mathbb{Z}_d$, where
\begin{equation}
    \label{eq:boolfunc}
    f(x) = \sum_{\substack{i_1,\ldots,i_k \in V \\ \{ i_1,\ldots,i_k \} \in E}} x_{i_1} \cdots x_{i_k}
\end{equation}
For graph states, $f(x)$ is quadratic, i.e.
\begin{equation}
    f(x) = \sum_{\substack{i_1,i_2 \in V \\ \{ i_1,i_2 \} \in E}} x_{i_1} x_{i_2}
\end{equation}
while for $f(x)$ cubic or higher, $\ket{H}$ is a hypergraph state. Therefore, for quadratic $f(x)$, one has a representation of stabilizer states, up to LC equivalence. For $f(x)$ cubic or higher, $\ket{H}$ represents hypergraph states which contain ``magic'' states. Examples of magic states are the $\mathrm{CCZ}$ state and Toffoli states, constructed from appropriate gates. Thus
\begin{equation}
    \mathrm{CCZ} \ket{x_1 x_2 x_3} = \omega^{x_1 x_2 x_3} \ket{x_1 x_2 x_3}
\end{equation}
with
\begin{equation}
    \ket{\mathrm{CCZ}} = \mathrm{CCZ} \ket{+ \otimes^3}
\end{equation}
as an example of a magic hypergraph state. Similarly
\begin{equation}
    \ket{\mathrm{Toff}} = \mathrm{Toff} \ket{+ \otimes^3}
\end{equation}
Explicitly,
\begin{align}
    \mathrm{Toff} \ket{i,j,k} = \ket{i,j,ij+k, \mod{d}}
\end{align}

\subsection{Level-rank duality}
The level-rank duality of the Pauli operators $X,\tilde{X},Z,\tilde{Z}$ in \ref{sec:2b} allows one to express the graph and hypergraph states \eqref{7}, \eqref{9}, \eqref{10},
\eqref{13}, \eqref{14} as the level-rank duals of graph and hypergraph states for $SU(2)_{2k}$ and for $SU(2)_{2k+1}/\mathbb{Z}_2$ ($k$ even). Since the hypergraph states contain magic states, there is a one-to-one map between such magic states and their level-rank duals.

It has been shown, using level-rank duality, that a universal topological quantum computer based on Chern-Simons theory for $SU(2)_2$ \cite{2000quant.ph..1108F} also implies an analogous universal quantum computer based on $SU(3)_2$ \cite{schnitzer2018levelrank}. However, this result depends on the level-rank duality of the Jones representation of the braid group, which differs from the duality discussed in this paper.

\acknowledgments
We thank Jonathan Harper and Isaac Cohen-Abbo for their help in preparing the manuscript.

\bibliographystyle{JHEP}
\bibliography{main}
\end{document}